Exploring new techniques for analyzing variability in white dwarf KIC 8626021

Thomas Huckans, Peter Stine

Department of Physics and Engineering, Bloomsburg University of Pennsylvania, 400 E 2nd St.,

Bloomsburg, PA 17815

Abstract

As is common with the collection of astronomical data, signals are frequently dominated by noise. However, when performing FTs of light curves, re-binning data can improve the signal-to-noise ratio (SNR) at lower frequencies. Using data collected from the Kepler space telescope, we sequentially re-binned data three times to investigate the SNR improvement of lower frequency (< 17 μHz) variability in white dwarf KIC 8626021. We found that the SNR at approximately 5.8 μHz greatly improved through this process, and we postulate that this frequency is linked to the rotation of KIC 8626021.

Introduction

First detected in 1862, white dwarfs long posed a mystery for early observers. When the companion to Sirius was detected, apparent contradictions concerning the mass, luminosities, and densities baffled astronomers. Lacking full understanding of atomic structures and the energy states of electrons, these early researchers believed white dwarfs too dense to exist. However, new discoveries at the turn of the 20th century explained the existence of these stars, and between the world wars white dwarfs were increasingly studied and modeled (Holberg, 2009).

As stars age, those that lack the mass to become neutron stars and black holes become white dwarf stars, representing 98% of the stars in our galaxy (Winget & Kepler, 2008). They are composed of a core of carbon and oxygen ions that slowly cools over billions of years, and the light emanating from these stars is a result of thermal energy. White dwarf stars are no longer supported against the force of gravity by fusion, so the stars collapse into an electron-degenerate state where the electrons in the carbon and oxygen atoms occupy the lowest energy levels. As two electrons cannot occupy the same quantum state, Pauli repulsion keeps white dwarfs from collapsing entirely.

For many years, accurate detection of light variability in white dwarfs was difficult due to a lack of adequate instruments. However, the launch of the Kepler space telescope in 2009 made capturing the light of distant stars much more efficient and effective (Basri et al., 2010). Kepler was initially developed with the intention of surveying our region of the Milky Way galaxy in order to find potentially habitable planets. The purpose of the mission was to identify key traits for such planets by determining the number of planets in habitable zones, the sizes and shapes of orbits, and the characteristics of the stars being orbited. Over the lifespan of its first mission, Kepler observed approximately $1.5 \times 10^5$ stars (Johnson, 2018), affording scientists excellent opportunities to research stellar variability. Due to the loss of a second reaction wheel in 2013, NASA developed the K2 mission, a way to prolong Kepler's assistance to astronomy and astrophysics.

Utilizing Kepler's ability to maintain three-dimensional control, NASA proceeded to use the telescope to collect photometry data of certain sections of our galaxy, although the number of targets was significantly reduced. In addition, the K2 mission was designed to be community-oriented, with the scientific community having an influence on the fields observed and serving as the analysts of the vast amounts of data being received (Howell et al., 2014). Although Kepler was deactivated in 2018, the data used in this paper came from observations during 2010 and 2012 of white dwarf KIC 8626021 and was obtained from the Kepler Asteroseismic Science Operations Center (KASOC).

The DBV white dwarf KIC 8626021 has an atmosphere rich in helium. Building upon previous studies, this research investigated novel techniques of analyzing variability in white dwarfs. The dwarf KIC 8626021 was chosen due to the large amount of preexisting research on the star, allowing for the validation of results using our methods. KIC 8626021 has an effective temperature of 29,700 K, log $g$ = 7.890, and mass of 0.56 M$_\odot$ (Córsico, 2020). Other research has found that this white dwarf is the DBV with the highest known temperature, and its helium layer is the thinnest (Bischoff-Kim et al., 2015). Despite the long-cadence light curve being too noisy to draw many conclusions, other FTs of short-cadence data have been performed to find variability in the dwarf. Analyses at high frequencies of KIC 8626021 yielded pulsations with frequencies of 4309.89 µHz, 5073.26 µHz, 3681.87 µHz, 3294.22 µHz and 2658.85 µHz (Østensen et al., 2011). These findings confirm the classification of the white dwarf as a V777 Herculis, although our research focuses on low frequencies using long-cadence data.

Methods

All data were downloaded from the KASOC database, and the long-cadence (data sampled approximately every thirty minutes) measurements of Corrected Flux (ppm) were analyzed. All computations were made in Wolfram Mathematica and Microsoft Excel, and FTs were performed in Mathematica. The re-binning process consisted of summing adjacent light curve data points in each quarter, therefore doubling the sampling interval from 0.5 hour to one hour, and then repeating this process on the data sample for a total of three times. In addition, a significant detection was defined as being 3$\sigma$ above the mean of the relative flux, and 0 on the graphs below represents this 3$\sigma$ cutoff. (Koch, D. G., 2010), (Wolfram Research, Inc., 2021). To find the SNR, we converted to decibels. Using these SNRs, we were able to easily identify improvement in signal strength.

Results

Figure 1 presents the lightcurves constructed for quarters seven (Q7) and thirteen (Q13), with corrected flux magnitude (ppm) plotted versus time (Julian days). Figure 2 presents the FTs of the first iteration and three successive re-bins for Q7, while Figure 3 presents the FTs of the same for Q13.

Tables 1 and 2 both show the hypothesized frequency corresponding to the rotation of KIC 8626021 that is found in the FTs of the first iteration and subsequent re-bins for Q7 and Q13. Tables 3 and 4 show all data values < 17 µHz found in the first iterations and re-bins of Q7 and Q13.

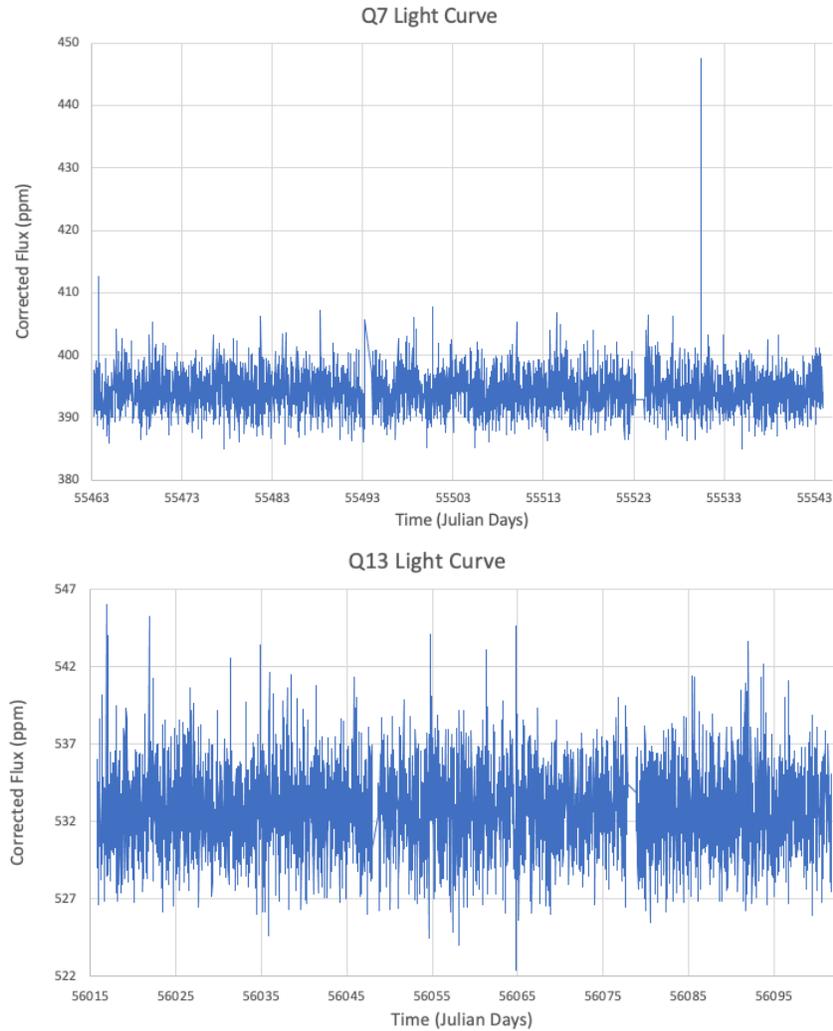

FIG. 1: Pictured top is the light curve constructed for Q7, below is the light curve for Q13. Q7 lasted from September 24 – December 13, 2010, and Q13 was from March 29 – June 23, 2012. Both graphs were constructed by plotting corrected flux magnitude (flux corrected for instrumental artifacts) versus time in Excel, and gaps in the data were filled in by interpolating between points. Q7 had forty-three interpolated points, and Q13 had sixty-six.

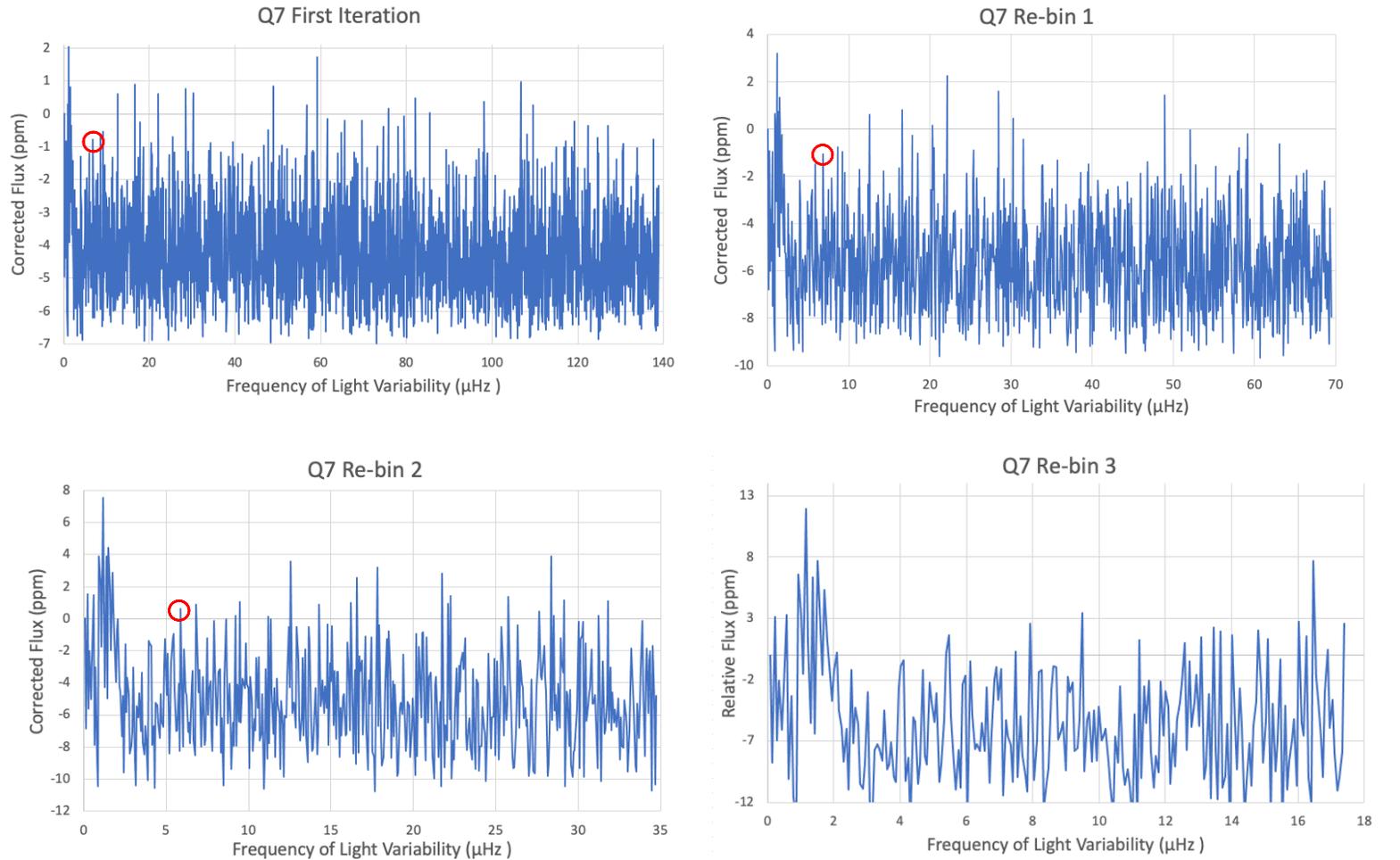

FIG. 2: The graphs show the initial FTs of Q7, and then the FTs of the three successive re-bins of the light curve data. The significant frequencies of 5.886 μHz and 5.889 μHz are circled. The disappearance of the frequency in the last FT is most likely a byproduct of the method, and the spurious frequency of 5.464 μHz in the last FT most probably represents an artifact of the re-binning process.

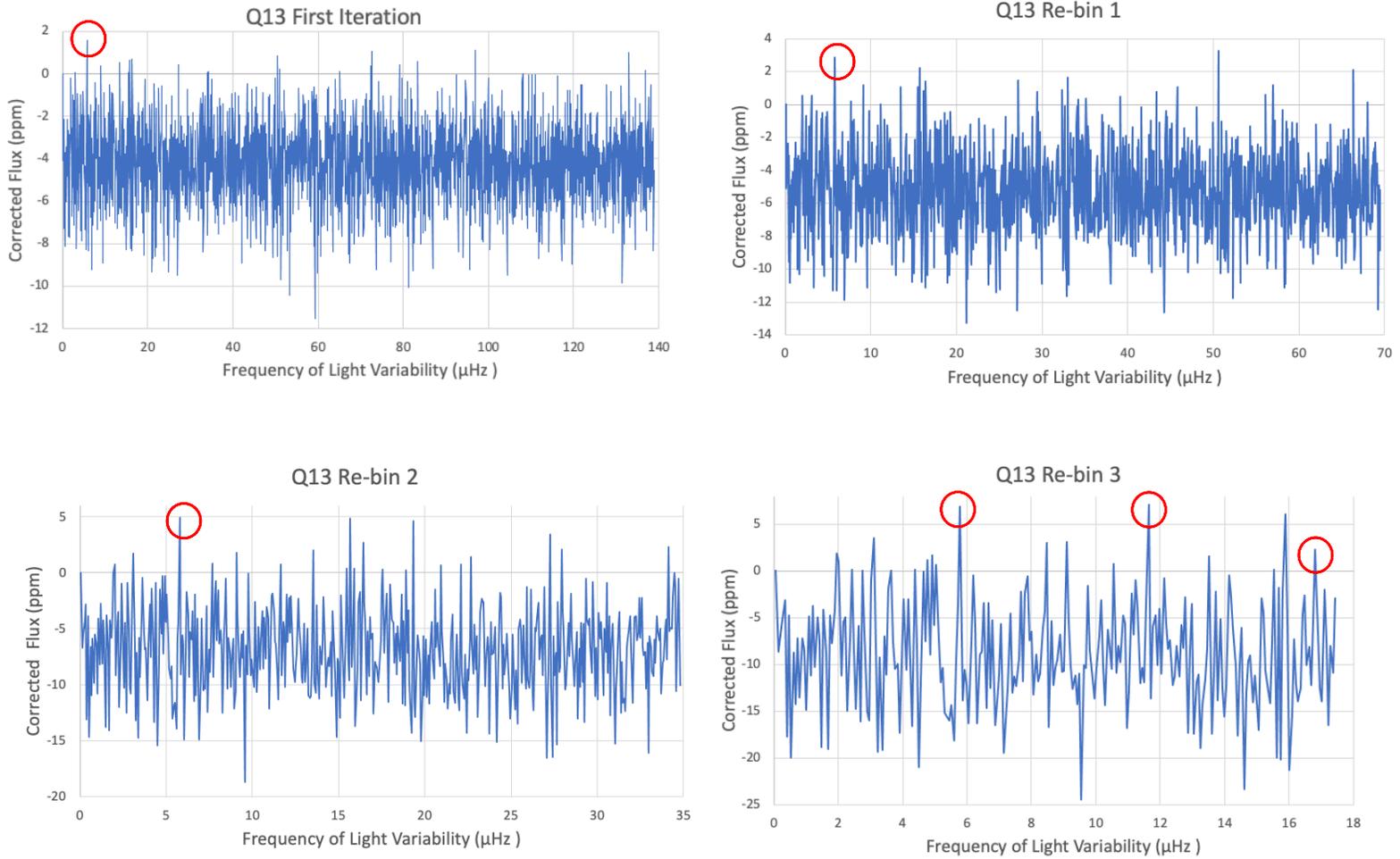

FIG. 3: The graphs show the initial FT of Q13, and then the FTs of the three successive re-bins of the light curve data. The significant frequencies of 5.784 µHz and 5.787 µHz are circled. In addition, in the third re-bin, the frequencies 11.641 µHz and 16.823 µHz rise above 3$\sigma$ and are nearly perfect integer multiples of 5.787 µHz. These harmonics are potentially indications of a starspot (Santos et al., 2017).

| Q7 Significant Data Points | Light Variability Frequency (µHz) | Corrected Flux Magnitude (ppm) | Period (days) | Signal-to-Noise (dB) |
|---|---|---|---|---|
| Q7 First Iteration | 5.886 | -1.198 | 1.966 | 9.9 |
| Q7 Re-bin 1 | 5.886 | -1.477 | 1.966 | 12.8 |
| Q7 Re-bin 2 | 5.889 | 0.597 | 1.965 | 19.2 |

TABLE I: The table displays the various frequencies collected from Q7 and the information found through calculations to find period and SNR. The frequency of 5.464 µHz is not included, and therefore was not used in any calculations determining the average period of rotation. The values under corrected flux magnitude are relative to our significant frequency cutoff of $3\sigma$, thus negative numbers are under the cutoff.

| Q13 Significant Data Points | Light Variability Frequency (µHz) | Corrected Flux Magnitude (ppm) | Period (days) | Signal-to-Noise (dB) |
|---|---|---|---|---|
| Q13 First Iteration | 5.784 | 1.555 | 2.001 | 15.6 |
| Q13 Re-bin 1 | 5.784 | 2.873 | 2.001 | 17.7 |
| Q13 Re-bin 2 | 5.787 | 4.938 | 2.000 | 22.6 |
| Q13 Re-bin 3 | 5.787 | 6.909 | 2.000 | 26.3 |
| Q13 Re-bin 3 | 11.641 | 7.073 | 0.994 | 26.4 |
| Q13 Re-bin 3 | 16.823 | 2.299 | 0.688 | 24.1 |

TABLE II: The table displays the various frequencies collected from Q13 and the information found through calculations to find period and SNR. The last two significant frequencies (11.641 µHz and 16.823 µHz) for Q13 Re-bin 3 represent potential harmonics, which are discussed in further detail in the Conclusions section of this paper. The values under corrected flux magnitude are relative to our significant frequency cutoff of $3\sigma$, thus negative numbers are under the cutoff.

| First Iteration (μHz) | First Re-bin (μHz) | Second Re-bin (μHz) | Third Re-bin (μHz) |
| --- | --- | --- | --- |
| 0.933 | 0.933 | 0.215 | 0.216 |
| 1.148 | 1.148 | 0.575 | 0.575 |
| 1.364 | 1.364 | 0.934 | 0.935 |
| 1.507 | 1.507 | 1.005 | 1.006 |
| 12.561 | 12.561 | 1.149 | 1.150 |
| 16.581 | 16.581 | 1.221 | 1.222 |
| | | 1.364 | 1.366 |
| | | 1.508 | 1.509 |
| | | 1.580 | 1.582 |
| | | 1.724 | 1.725 |
| | | 1.795 | 1.797 |
| | | 5.889 | 2.085 |
| | | 6.822 | 5.392 |
| | | 9.192 | 5.464 |
| | | 9.479 | 7.476 |
| | | 11.203 | 9.489 |
| | | 12.568 | 11.215 |
| | | 14.291 | 12.581 |
| | | 16.230 | 13.084 |
| | | 16.589 | 13.443 |
| | | | 13.659 |
| | | | 14.018 |
| | | | 14.809 |
| | | | 15.097 |
| | | | 16.031 |
| | | | 16.463 |
| | | | 16.894 |

TABLE III: The table displays all frequencies of Q7 that had a corrected flux magnitude (ppm) above the cutoff of $3\sigma$. The minor shifting of significant frequencies between re-bins is a by-product of the method, and we calculated for such errors when finding our average.

| First Iteration (µHz) | First Re-bin (µHz) | Second Re-bin (µHz) | Third Re-bin (µHz) |
| --- | --- | --- | --- |
| 3.094 | 2.018 | 2.019 | 1.951 |
| 5.784 | 3.094 | 3.095 | 2.019 |
| 9.080 | 5.784 | 5.787 | 2.442 |
| 13.519 | 7.667 | 7.671 | 2.759 |
| 15.671 | 9.080 | 9.084 | 3.095 |
| 16.209 | 11.165 | 11.641 | 3.634 |
| 16.411 | 13.519 | 13.526 | 4.374 |
| | 15.469 | 15.477 | 4.778 |
| | 15.671 | 15.679 | 4.912 |
| | 16.209 | 15.881 | 5.047 |
| | 16.411 | 16.419 | 5.787 |
| | | | 8.479 |
| | | | 9.084 |
| | | | 10.565 |
| | | | 11.641 |
| | | | 13.526 |
| | | | 15.544 |
| | | | 15.881 |
| | | | 16.823 |

TABLE IV: The table displays all frequencies of Q13 that had a corrected flux magnitude (ppm) above the cutoff of 3$\sigma$. The minor shifting of significant frequencies between re-bins is a by-product of the method, and we calculated for such errors when finding our average.

## Conclusions

As our research used the long-cadence data from Kepler, much of the high-frequency variability due to gravitational wave pulsations is lost. However, this presents an opportunity to verify our results with the work of research groups that analyzed short-cadence data. With the data analyzed, the lower frequencies between 5-6 µHz emerged. After finding the average of the periods and accounting for a 1$\sigma$ margin of error, our research hypothesizes that the rotation period of KIC 8626021 is 1.99 ± 0.02 days. Other short-cadence research has found the rotation period to be 1.8 ± 0.4 days, by analyzing the structures of independent modes (Bischoff-Kim et al., 2015). Other calculated periods of rotation have been ≈ 1.7 days (Østensen et al., 2011), and these periods indicate that the more precise significant period identified through our re-binning relates to the rotation of the white dwarf.

Through the re-binning process, the SNR clearly improves for both quarters, and for Q7 it improves by approximately 1.3 dB, except for the last data re-bin. In the last re-bin, the previous significant frequency disappears, which becomes increasingly likely after successive re-binning processes. The frequency 5.464 µHz rises as another significant frequency; however, we believe that this new frequency is simply an artifact of the re-binning process. In Q13, we saw SNR improvement ranging from 1.1 dB to 1.3 dB.

Through the re-binning process, more lines, or significant frequencies, appeared above the 3$\sigma$ cutoff, particularly at lower frequencies. These findings suggest that as an alternative to short-cadence analysis, the re-binning process of long-cadence data can be used to identify significant lower frequencies in white dwarfs. The methods we used are also simple and replicable, which allows even those with less experience to quickly analyze the large amounts of data being collected by orbiting telescopes, such as the currently active TESS (Transiting Exoplanet Survey Satellite) telescope.

The presence of possible harmonics in the third re-bin of Q13 also indicates the possible presence of a previously unseen starspot in KIC 8626021 caused by magnetic activity. These spots are darker, cooler, and modulate stellar light curves, and with confirmation of its existence, the harmonic frequencies can be used to calculate the spot's rotation rate, size, latitude, and contrast (Santos et al., 2017). Using the process of re-binning, a starspot signal, previously dominated by noise, may have been discovered.

Acknowledgments

We wish to thank Bloomsburg University of Pennsylvania for its continued support of our research.

This paper includes data collected by the Kepler mission and obtained from the MAST data archive at the Space Telescope Science Institute (STScI). Funding for the Kepler mission is provided by the NASA Science Mission Directorate. STScI is operated by the Association of Universities for Research in Astronomy, Inc., under NASA contract NAS 5–26555.